\documentclass[epj]{svjour}
\usepackage{epsfig}
\usepackage{amsmath}
\usepackage{amsfonts}
\usepackage{amssymb}
\usepackage{times}

\newcommand{\be}{\begin{equation}}
\newcommand{\ee}{\end{equation}}
\newcommand{\bea}{\begin{eqnarray}}
\newcommand{\eea}{\end{eqnarray}}
\newcommand{\ba}{\begin{array}}
\newcommand{\ea}{\end{array}}

\begin{document}
\title{Monitoring noise-resonant effects in cancer growth influenced by external fluctuations and periodic treatment}

\author{Alessandro Fiasconaro \inst{1}\inst{2}\inst{3} \thanks{email: afiasconaro@gip.dft.unipa.it} \and Anna Ochab--Marcinek\inst{2}\inst{4}  \and Bernardo Spagnolo \inst{3} \and Ewa Gudowska--Nowak\inst{1}\inst{2}}

\institute{Mark Kac Complex Systems Research Center,
Jagellonian University, Reymonta 4, 30-059 Krak\'ow,
Poland.
 \and
 Marian~Smoluchowski Institute of Physics, Jagellonian
University, Reymonta 4, 30-059 Krak\'ow, Poland.
 \and
Dipartimento di Fisica e Tecnologie Relative and CNISM,
Group of Interdisciplinary Physics, Universit\`a di
Palermo, Viale delle Scienze, I-90128 Palermo, Italy.
 \and
 Institut f\"ur Physik,
Universit\"at Augsburg, Universit\"atsstra\ss{}e 1, 86135
Augsburg, Germany. }

\date{Received: date / Revised version: date} 
\date{\today}

\abstract{In the paper we investigate a mathematical model
describing the growth of tumor in the presence of immune
response of a host organism. The dynamics of tumor and
immune cells is based on the generic Michaelis-Menten
kinetics depicting interaction and competition between the
tumor and the immune system. The appropriate
phenomenological equation modeling cell-mediated immune
surveillance against cancer is of the predator-prey form
and exhibits bistability within a given choice of the immune response-related parameters. Under the influence of weak external
fluctuations, the model may be analyzed in terms of a
stochastic differential equation bearing the form of an
overdamped Langevin-like dynamics in the external
quasi-potential represented by a double well. We analyze
properties of the system within the range of parameters for
which the potential wells are of the same depth and when
the additional perturbation, modeling a periodic treatment,
is insufficient to overcome the barrier height and to cause
cancer extinction. In this case the presence of a small
amount of noise can positively enhance the treatment,
driving the system to a state of tumor extinction. On the other hand, however, the same
noise can give rise to return effects up to a stochastic
resonance behavior. This observation provides a
quantitative analysis of mechanisms responsible for
optimization of periodic tumor therapy in the presence of
spontaneous external noise. Studying the behavior of the extinction time as a function of the treatment frequency, we have also found the typical resonant activation effect: For a certain frequency of the treatment, there exists a minimum extinction time.
\PACS{
       {05.40.-a} {} \and
       {87.17.Aa} {} \and
       {87.15.Aa} {}
     }
} 

\authorrunning {Alessandro Fiasconaro et al.}
\titlerunning {Monitoring noise-resonant effects in cancer growth...}
 \maketitle

\section{Introduction}

Although cancer is a leading cause of death in the world,
it is still little known about the mechanisms of its growth
and destruction. Surgery, chemo- and radio-therapies play
key roles in treatment, but in many cases they do not
represent a cure. Even when patients experience tumor
regression, later relapse can occur. The need to address
more successful treatment strategies is clear. Currently,
efforts are made to investigate, among others, the methods
of adoptive cellular immunotherapy
\cite{Rosenberg_Science,Dillman}.
These methods of tumor treatment are based on the use of
the injection of cultured immune cells, which have
anti-tumor reactivity, into the tumor-bearing host.
Therefore, a detailed theoretical study on the mechanisms
of interaction between tumor tissue and immune system is
necessary for planning efficient strategies of treatment
\cite{Thorn_Henney,Moy_Holmes_Golub,Kirschner_Panetta,Chaplain}.
The immune response against the
antigens generated by certain tumors, may be mediated by so called effector cells
such as T-lymphocytes, macrophages or natural killer cells.
The process of damage to tumor proceeds via infiltration of
the latter by the specialized cells which subsequently
develop a cytotoxic activity against the cancer
cell-population. The series of cytotoxic reactions between
the immune cells and the tumor tissue may be considered to
be well approximated
\cite{Thorn_Henney,Moy_Holmes_Golub,Kirschner_Panetta,Chaplain,GARAY,Le_Fever}
by a saturating, enzymatic-like process whose time
evolution equations are similar to the standard
Michaelis-Menten kinetics: The development of tumor tissue
and its reaction to immune response can be described in
terms of a predator-prey model
\cite{Chaplain,GARAY,Le_Fever,Ochab_PA,AlePRE06,Ochab_APP2,Spagnolo_app07,Lefever_Garay,Lefever_Horsthemke,Prigogine_Lefever,Horsthemke_Lefever,Ebeling,Gudowska1,Gudowska2,Michor,Molski}.
The population of tumor cells plays the role of "preys",
and immune cells act as "predators". The activity of the
predator in a certain territory, or, in this case, the
activity of immune cells in tissue, resemble the mode of
action of enzymes or catalysts in a chemical reaction,
where  the enzymes transform substrates in a continuous
manner without destroying themselves. The constant immune
cell population is assumed to act in a similar way, binding
the tumor cells and releasing them unable to replicate.

The network of tumor-immune system interactions is
subjected to random fluctuations. The growth rate of tumor
tissue is affected by many environmental factors, e.g. the
degree of vascularization of tissues, the supply of oxygen,
the supply of nutrients, the immunological state of the
host, chemical agents, temperature, radiations, etc. As a
result of this complexity, it is unavoidable that in the
course of time the parameters of the system undergo random
variations which give them a stochastic character
\cite{GARAY,Lefever_Horsthemke,Gudowska1,Gudowska2,Michor}.

In our previous works the effect of noise in the cancer dynamics has
been studied both analytically and numerically finding in
its evolution the well known phenomena of resonant
activation (RA) \cite{ra} and the noise enhanced stability (NES)
\cite{Ochab_PA,AlePRE06,Ochab_APP2,Spagnolo_app07,Agudov_301}.

The aim of this study is to explore the dependence between
the intensity of external fluctuations and the transition
time from the state of a stable tumor to the state of its
extinction under the influence of a moderate periodic
treatment. Low intensity of therapy presents the advantage
of minimizing side effects of treatment. The latter, if
occasionally combined with spontaneous environmental
fluctuations, can also cause the re-occurrence of cancer
bearing a similarity to a stochastic resonance (SR) effect.
Accordingly, our studies aim to understand a competition
between the probability of the extinction and the
reappearance of the cancer cells in the system, and to
analyze optimal conditions of its control.

\section{The Model System}

The formulation of the model describing the growth of
cancerous tissue attacked by immune cytotoxic cells is
based on a reaction scheme
\cite{Chaplain,GARAY,Lefever_Garay,Lefever_Horsthemke,Prigogine_Lefever,Horsthemke_Lefever,Ebeling,Gudowska1,Gudowska2}
representative of the catalytic Michaelis-Menten scenario:
 \bea
  X &\longrightarrow& \!\!\!\!\!\!\!\!\!\!^{\lambda} \quad 2X  \nonumber\\
 X\ +\ Y\  \longrightarrow&
 \!\!\!\!\!\!\!\!\!\! ^{k_{1}}\ \ \ \ Z\ &\longrightarrow
 \!\!\!\!\!\!\!\!\! ^{k_{2}}\ \ \ \  Y\ +\ P\ .
 \eea
Here cancer cells $X$ are assumed to proliferate
spontaneously at a rate $\lambda$ whereas their local
interactions with cytotoxic cells $Y$ are modeled by a
simplified kinetics with $k_1$ standing for the rate of
binding of immune cells to the complex $Z$ which
subsequently dissociates at a rate $k_2$. The dissociation
results in a product $P$ representing dead or
non-replicating tumor cells. In the limit case, when the
production of $X$-type cells inhibited by a hyperbolic
activation is the slowest process under consideration, and
by assuming a conserved number of immune cells
$Y+Z=E=const$, the resulting kinetics can be recast in the
form of the first order differential equation. This can be
interpreted as an overdamped equation of motion in a
pseudo-potential $U(x)$ (see
\cite{Lefever_Garay,Lefever_Horsthemke,Prigogine_Lefever} for the details)
\be
 \frac{dx}{dt}=-\frac{dU(x)}{dx}
\ee where
 \bea
  \label{eq:pot}
  U(x)=-\frac{x^2}{2}+\frac{\theta x^3}{3}+\beta x -\beta
  \ln(x+1),
  \label{Ux}
 \eea
$x$ stands for the normalized molecular density of
cancerous cells with respect to the maximum tissue capacity
and the following scaling relations for the kinetic
parameters have been used
 \bea
 x=\frac{k_{1}x}{k_{2}}, \,\,\,\,\,   \theta=\frac{k_{2}}{k_{1}}, \,\,\,\,\,
\beta=\frac{k_{1}E}{\lambda}, \,\,\,\,\, t=\lambda t.
\label{eq:scaling}
 \eea
\begin{figure}[t]
  \epsfig{figure=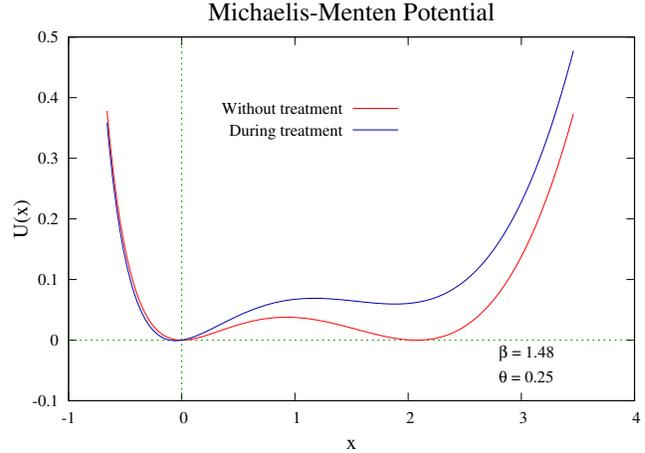, angle=-90, width=8.8cm}
 \caption{Pseudo-potential $U(x)$ (Eq.
\ref{Ux}) with parameters: $\beta=1.48, \theta=0.25$
that assure bistability with an almost symmetric character
(equal depths) of both potential wells. The minima are
located  at $x_1 = 0$ and $x_3 \approx 2.08$, separated by
a barrier at $x_2 \approx 0.925$. During the therapy (see
the text below), this profile tilts slightly due to the
driving $d(t)=-A[1-\Theta(\cos(2\pi\nu t))]$ which causes
alternating lowering and raising of the barrier. The cyclic
variation of the potential profile is depicted here for the
intensity $A=0.03$ and period $T=1/\nu=1000$ in units
chosen accordingly to the scaling (See
Eq.~(\ref{eq:scaling})). The minimum on the right side is still present during the treatment. Note a
non-physical character of the left-hand side profile beyond
$x_1=0$. The vertical draft line indicate the reflecting
boundary for $x=x_1$ in order to avoid this possibility on
our calculations.}
 \label{fig:pot}
\end{figure}
Typical experimental values of the parameters are
\cite{GARAY,Prigogine_Lefever,Horsthemke_Lefever}: $k_1 =
0.1-18 \; day^{-1}$, $k_2 = 0.2-18 \; day^{-1}$, $\lambda =
0.2-1.5 \;day^{-1}$. In the above deterministic model, weak
fluctuations, that is temperature variations or changes in
local concentrations of biochemical agents, can be
incorporated by including additional source of stochastic
fluxes represented by an additive noise $\xi(t)$
\be
 \frac{dx}{dt}=-\frac{dU(x)}{dx}+ \xi(t)
 \label{lang}
\ee
which is assumed to be uncorrelated and Gaussian
distributed with zero mean: $\langle \xi(t)\xi(t') \rangle
= D \delta(t-t')$, and $\langle \xi(t) \rangle = 0$, with
the constraint $x\geq 0$. In the above formulation the
parameter $D^{1/2}$ stands for the intensity of the fluctuations.
 The resulting stochastic differential
equation Eq.(\ref{lang}) is Langevin-type.
Alternatively, the dynamics of the system described by the
Langevin equation may be characterized by determining the
probability density $p(x,t)$ of $x$ which carries
information about the instantaneous state of the system and
fulfills the Fokker-Planck equation
 \bea
  \frac{\partial}{\partial t} p(x,t)=\frac{\partial}{\partial x} \left[\frac{\partial U(x)}{\partial x} p(x,t) \right] +\frac{D}{2} \frac{\partial^2 p(x,t)}{\partial x^2} .
  \label{fpe}
\eea
The estimation of the fluctuation in cancer dynamics can be obtained by analyzing the cancer growth trajectories. An example of paper
showing the time series for different patients is reported in
\cite{Shirato}, where we observe fluctuations in agreement with the
range here investigated.

The potential profile Eq.~(\ref{Ux}) exhibits three
equilibrium states of the system (see Fig.~\ref{fig:pot}):
\bea x_1&=&0, \\
 x_2&=&\frac{1-\theta+\sqrt{(1+\theta)^{2}-4\beta
   \theta}}{2\theta}, \\
 x_3&=&\frac{1-\theta-\sqrt{(1+\theta)^{2}-4\beta
   \theta}}{2\theta}.
 \eea
The essential feature captured by the model is, for a
constant parameter $\theta$, the $\beta$-dependent
bi-stability (\cite{GARAY,Ochab_PA,AlePRE06,Ochab_APP2,Spagnolo_app07,Lefever_Garay,Lefever_Horsthemke,Prigogine_Lefever,Gudowska1,Gudowska2}) of stationary (long-time) solutions. In a certain range of values of the above parameters the model
possesses two stable states: the state of extinction, where
no tumor cells are present, and the state of stable tumor,
where its density does not increase but stays at a certain
constant level. The bistable behavior of the system can be also observed by
examining the stationary (time-independent) solution
$p_{eq}=const\times e^{-U(x)/D}$ to Eq.(\ref{fpe}) which,
for small value of the noise intensity $D^{1/2}$,  is
sharply peaked around the minima of the potential $U(x)$.

In the present work we are interested in situations where
the stationarity of the system is broken due to coupling to
an external driving representing a controllable therapy.
Here, the influence of therapeutic factors is studied by
considering a periodic treatment (chemo- or radiation-
therapy) whose action, at the level of the kinetic equation
can be modelled by a contribution:
\bea
  d(t) =  -A [1- \Theta(cos(2 \pi \nu t))],
  \label{eq.dose}
 \eea
to the RHS of Eq. \ref{lang}. This term results in a
periodic decrease of $X$ cell concentration. The
$\Theta$-symbol stands here for the Heaviside function
reflecting the on-off switch of the cyclic treatment
performed with the intensity $A$ and frequency $\nu$.
Accordingly, the $d(t)$ term leads to a modulation of the
pseudo-potential whose time-dependent part is given by:
 \bea
  U_d(x,t) =  Ax [1- \Theta(cos(2 \pi \nu t))].
  \label{eq.dosepot}
 \eea
The overall kinetic equation describing the evolution of
the cancer cells population is then of the form
 \bea
  \label{eq.all}
  \frac{dx}{dt} & = &-\frac{d(U(x)+U_d(x,t))}{dx}+\xi(t) \\
   &=& x(1-\theta x) - \frac{\beta x}{x+1} -A [1- \Theta(cos(2 \pi \nu t))] + \xi(t) \nonumber
 \eea
For the purpose of these studies, we consider the situation
in which both wells of $U(x)$ are of equal depth, using the parameters $\theta=0.25$, and
$\beta=1.48$ (See Fig.~\ref{fig:pot}).  Since
the cell number $X$ cannot be negative, we investigate
numerically the properties of the system corresponding to
the Langevin equation (\ref{eq.all}) under the constraint
$x\geq 0$.
In the absence of noise, the existence of the potential
barrier prevents the trajectories (solutions to
Eq.(\ref{eq.all})) from switching between potential wells,
which correspond to well defined cancer states. If noise is
present but there is no periodic forcing ($A=0$), a
spontaneous tumor extinction or occurrence become possible:
solutions to the Langevin equation will cross the potential
barrier at random times whose expectation value is given by
an inverse of Kramers rate, i.e. $\tau\propto e^{\Delta
U(x)/D}$ where $\Delta U(x)$ stands for the height of the
barrier. Under the action of periodic forcing
$-d(U_d(x,t))/dx$, representing action of irradiation or
chemical therapy, the overall potential profile changes and the barrier will still be crossed at random
times but with a preference for the instants of minimal
barrier height. In particular, if the noise intensity is
large enough compared to the barrier height $\Delta U$,
transitions between the wells become likely to occur twice
per period. This phenomenon, addressed in literature
\cite{hanggi,mcnamara} as a stochastic resonance (SR), is
a generic phenomenon manifested in nonlinear
systems where a weak signal can be amplified and optimized
by the presence of noise. Obviously, if discussed in the
context of periodic therapy administration, the SR effect
(multiple synchronized re-crossings of the barrier) should
be avoided as leading to possible recurrence and regrowth
of tumor.

The aim of our study is therefore to investigate the role
of weak random contributions to deterministic dynamics of
the tumor growth model in the case when the intensity
(amplitude) of cyclical treatment is subthreshold, i.e. not
sufficient to induce cancer extinction. A modeled action
of irradiation/chemical treatment is studied based on the
one-dimensional kinetic equation (\ref{eq.all}), expressing
growth of neoplastic cells population restricted to a
saturation level and subject to the immunological
surveillance mechanism of a host. Neither a particular
histological type of tumor, nor other factors determining
cancer response to fractionated treatment (like
redistribution of cells in various phases of cell cycle
\cite{Andersen}) are specified. Within this simplified
description, our main purpose is to analyze the
effectiveness of a treatment intended to cause the
extinction of a tumor which is already present in a tissue
and whose size is described by the density of cancer cell
population $x_3$.
 \begin{figure}[hbt]
 \epsfig{figure=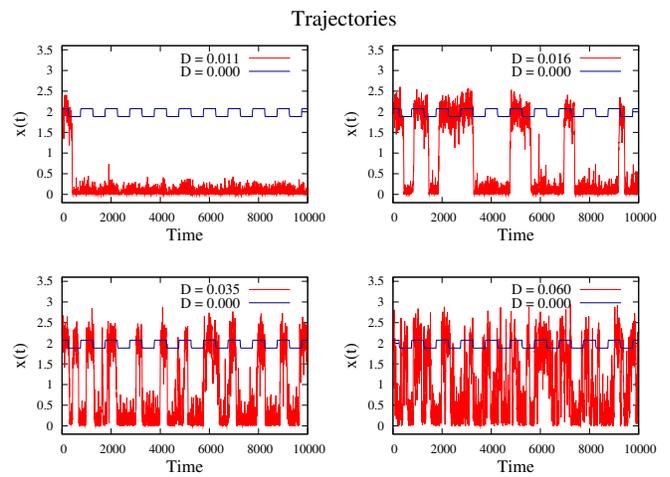, angle=-90, width=8.8cm}
 \caption{\small Deterministic ($D=0$) and
stochastic ($D>0$) trajectories of cancer evolution with
the same parameters of Fig.\ref{fig:pot}. For $D=0$, the trajectory is localized in the cancer state at the minimum of the well. In contrast, a low level of noise is instead
able to induce the extinction acts ($D=0.011$). For
increasing intensities of noise, the reappearance of cancer
is detected ($D=0.016$). In turn, for noise intensity
close to the value $D=0.035$ a stochastic resonance effect
becomes visible.}
 \label{fig:xt}
 \end{figure}

\section{Results}
The extinction and the
reappearance of the tumor cells are studied by the
statistics of both the first extinction time (FET) and the
first return time (FRT). We define FET as the time needed
by a trajectory $x(t)$ starting from the stable state
$x(0)=x_3$, to reach the value $x=0$ for the first time.
Similarly, the FRT is the time needed by a trajectory
$x(t)$ to return to the initial state $x_3$ for the first
time, after having reached the value $x=0$. To avoid the
unphysical situation of trajectories crossing the state
$x=0$, we place there a reflecting boundary when analyzing
the FRT, while an absorbing boundary is present in the
analysis of FET. Numerical evaluation of typical
trajectories $x(t)$ established for the "dose"-intensity of
a treatment $A=0.03$ is depicted in Fig.~\ref{fig:xt}.
Without addition of noise ($D=0$) the concentration of
cancer cells remains close to the $x_3$ level and the
$x(t)$ trajectories are localized within the right
potential well (cf. Fig.~\ref{fig:pot}). In turn, a small
amount of noise added to the system can give rise to cancer
extinction whose occurrence is displayed  for $D=0.011$.

The realistic values of the noise intensity used in this work, or the corresponding standard deviation $\sigma = \sqrt{D} \in [0.005,0.32]$, can be estimated using the scaling of the Eq. \ref{eq:scaling} and the experimental values of parameters (see the previous Section), obtaining $\sigma_{Real} \approx \lambda \sqrt{D} \in [0.014, 0.47]$~1/day, or, in percentage of the maximum tumor cell density, $\sigma_{Real} \in [1.4\%, 47\%]$.

\begin{figure}[hpbt]
 \epsfig{figure=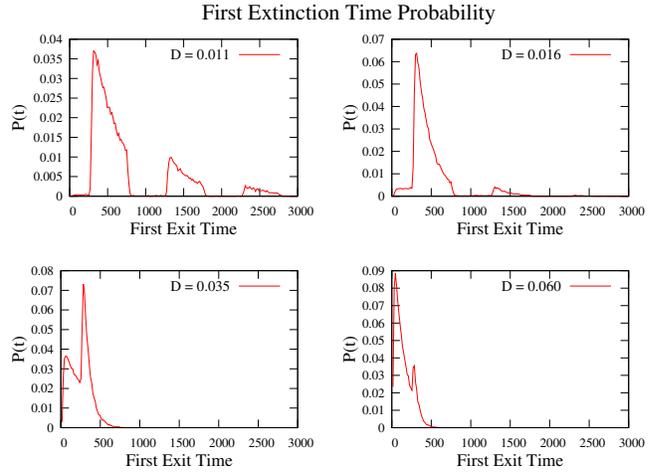, angle=-90, width=8.8cm}
 \caption{\label{fig:fetp} \small
Probability distribution of extinction times. At very low
level of the noise intensity, extinction events are
significantly probable only during the treatment periods
whereas no extinction is detected in the "treatment-off"
periods. By increasing noise intensities the extinction
events occur also in the no-treatment intervals.
Consequently, differences between the response during
"treatment-on" and "treatment-off" intervals decrease
significantly.}
\end{figure}

The statistics of such occurrences observed on the ensemble
of  $N=5\times 10^{4}$ realizations $x(t)$ is shown in Fig.~\ref{fig:fetp} for different noise intensities. At
values of $D$ of order of $0.011$ separate peaks in histograms of FETs are visible, suggesting that the extinction events
(passages from $x_3$ to $x=0$ state) are possible only
during the "treatment-on" session, beginning at $t=T/4$ up
to $t=3T/4$, and repeated recursively in a period $T =
1/\nu = 1,000$ (in unit of $1/\lambda$).
According to the experimental data mentioned in the previous Section, the period of treatment used in our simulations lies in the range (using $T_{real} = T / \lambda$~[days]), $T_{Real} \in [666,5000]$~days. In literature cancer recurrence has been observed from $9$ up to $17$ months after a chemical treatment time of at least $10$ months \cite{Papatsorisa}, so the low values of that range seem to be the most realistic.

\begin{figure}[hptb]
 \epsfig{figure=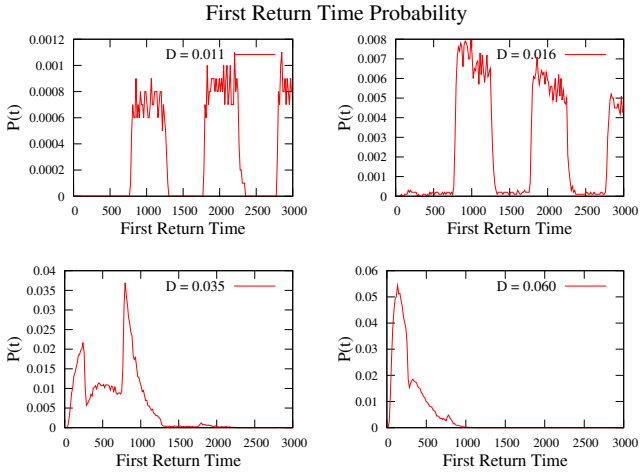, angle=-90, width=8.8cm}
 \caption{\label{fig:frtp} \small
Probability density of return times. For a low level of the
noise intensity the first return event is observed during
the no-treatment time interval. By increasing the noise
intensity the return events happens also during the
treatment (cf.Fig.\ref{fig:fetp}). At high noise intensity,
the first return time- and the first exit time
distributions tend to become similar each other regardless
of the state of the treatment.}
\end{figure}

 The maximum temporal window explored is  $t_{max}=10T$. The
distinction between the two intervals ("on" and
"off" treatment response) becomes blurred at higher noise
intensities as visible in Fig.~\ref{fig:fetp}. Moreover, by
increasing the noise, the extinction events occur at
earlier times the maximum of the probability $P(t)$
shifts towards shorter first exit times). This means that
the "energy supply" pumped into the system by the noise
contributes positively to the response of the tumor cell
population towards extinction even in the no-treatment
intervals (cf. Fig.~\ref{fig:fetp}, left-bottom panel,
$D=0.035$).
To elucidate the role of the spontaneous fluctuations
around the deterministic trajectories, we have analyzed the
statistics of cancer recurrence as addressed in
Fig.~\ref{fig:frtp}. Cancer regrowth events occur first
during the no-treatment time-interval for low level of
noise intensity ($D=0.011$) and, by increasing the
noise-intensity,  may be also detected during the
"treatment-on" intervals. This apparent competition effect
between the extinction and recurrence events will be
addressed further by analyzing the ratio between the mean
first exit time (MFET) and the mean return time (MFRT).

At low value of the noise $D =0.011$, the return probability is more
than one order of magnitude lower than the extinction
probability estimated for a given time window (compare
Figs.~\ref{fig:fetp} and \ref{fig:frtp}). This is due to
the fact that the starting state for a temporal evolution
of the $X$-cells concentration is $x_3$. Consequently, the
(first) returning trajectory $x(t)$ needs more time to
overpass the barrier, to hit the zero-concentration state
and to continue further in a reverse direction, than an
exiting trajectory which stops at $x_1$. In addition, the
potential barrier, in half a period, is higher in the return motion than in
the exit one, in which the treatment facilitates the
escape. On the other hand, at sufficiently high noise
intensity, the times to cross the barrier in forward
or backward direction are approximately equal (see the bottom
right panel of Fig.\ref{fig:xt}).

\begin{figure}[hpbt]
   \epsfig{figure=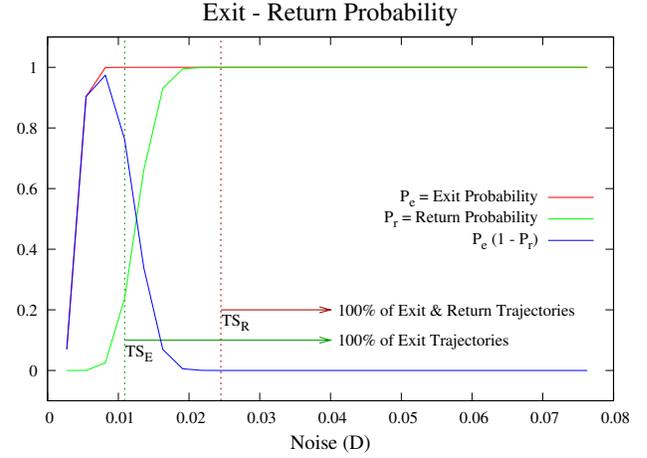, angle=-90, width=8.8cm}
\caption{\small Probability of the exit and return
occurrence as a function of the noise intensity. The 'exit without return'
probability $P_e (1-P_r)$ has a nonmonotonic behavior with
an optimal value of the noise intensity ($D \approx
0.008$) not far from $D_{opt}$ shown in the text.
 \label{fig:stat}}
\end{figure}

The total probability of the exit ($P_e$) and return ($P_r$)
occurrences as a function of the noise intensity calculated
as the ratio of the number of exit/return trajectories
normalized over the total number of experiments in the observation time window, is shown in
Fig.~\ref{fig:stat}. We note that for $D>0.011$ all the
trajectories leave the right well of the potential giving
extinction of the cancer. For $D>0.024$ all the
trajectories experience the return back to the well after
having reached the left minimum at $x_1$. In the
intermediate noise region ($0.011<D<0.024$) all the
trajectories give extinction but only a certain percentage
are able to return into the right well of the potential.
The 'exit without return'
probability $P_e (1-P_r)$ has a nonmonotonic behavior with
an optimal value of the noise intensity ($D \approx
0.008$) not far from $D_{opt}$ shown below.
The threshold lines $TS_E$ and $TS_R$ represent this
limiting values and will be indicated as reference lines in
the following plots.

\begin{figure}[hpbt]
   \epsfig{figure=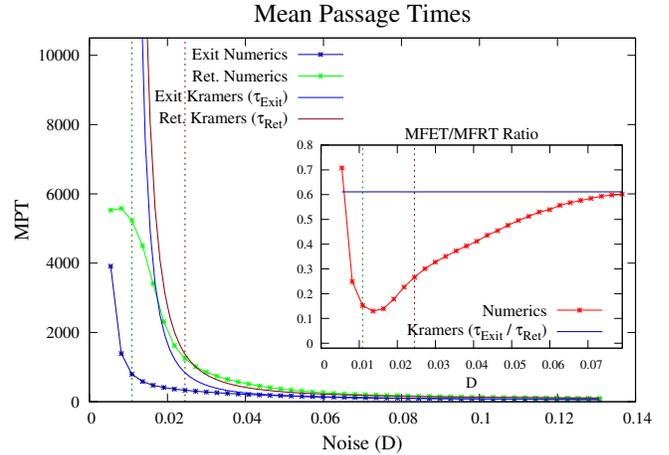, angle=-90, width=8.8cm}
\caption{\small Mean first exit time and mean first return
time as a function of the noise intensity. The two averages
have been made using the actual number of trajectories
having reached the related boundary ($x_1$ for the exit
case, $x_1$ and then $x_3$ for the return one). The
\emph{ratio} between both measures (displayed in the inset)
exhibits a minimum representing the best stochastic
compromise between effectiveness of the therapy and the
risk of cancer recurrence. It is worth to note that in a
very small region of noise intensity ($D \in
[0.013,0.015]$) we find both the minimum value for the
ratio of the MFET/MFRT and of the related width of the
distributions (See Fig.\ref{fig:mpt2}).}
 \label{fig:mpt}
\end{figure}

 The above described observations are well reflected in
the character of the mean first exit time (MFET) and the
mean first return time (MFRT), as derived from the related
probabilities $P(t)$. Fig.~\ref{fig:mpt} displays both mean
times as functions of the noise intensity $D$. At very low
noise intensities, an almost deterministic character of the
trajectories is mirrored in very high values of the MFRT
and MFET, respectively.
By increasing noise intensity, both MFET and MFRT decay
sharply and MFET tends to become proportional to MFRT
at a value of $D\approx 0.1$. A moderate noise intensity makes the
transitions between the wells more likely to occur and thus shortens the
average extinction time of cancer. On the other hand, at
higher noise intensities, duration of returning
trajectories is (roughly) twice longer than duration of
exiting ones.
 This behavior is consistent with the results
as predicted by inspection of $P(t)$ histograms: clearly,
at sufficiently high noise intensities distributions of
first exit and first return times become similar with some
asymmetry in shape related to the asymmetry of the
potential and to the constraints of motion imposed by the
boundary condition. Consequently, the ratio MFET/MFRT is slightly higher than one half, as indicated in
Fig.\ref{fig:mpt}.
 The two averages have been made using
the actual number of trajectories having reached the
related boundary ($x_1$ for the exit case, $x_1$ and then
$x_3$ for the return one), excluding the trapped
trajectories from the related calculations. The behavior of
the MFET and MFRT response, which is different for low
noise intensities, but similar for high noise intensities
(where the two means have the same trend and their values
tend to become proportional), accounts for the
non-monotonic behavior of the MFET/MFRT ratio as observed
in Fig.~\ref{fig:mpt}. In fact, for the chosen set of
parameters, the minimum of MFET/MFRT is observed for
$D_{opt}\approx 0.013$ which is the optimum
noise-intensity value giving the best compromise between
noise induced extinction of the cancer cells and their
reappearance.

\begin{figure}[hpbt]
   \epsfig{figure=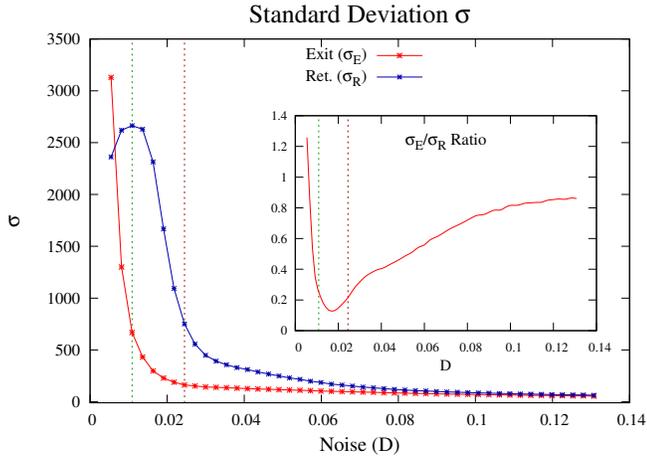, angle=-90, width=8.8cm}
\caption{\small Standard deviation derived from the
probability densities of first exit and first return times.
Similarly to the situation observed for the means, the
ratio $\sigma_{E}/\sigma_{R}$ attains a minimum within the
range of investigated noise intensities, very close to the
MFET/MFRT minimum at $D_{opt}$ (See Fig. \ref{fig:mpt}).}
\label{fig:mpt2}
\end{figure}

\begin{figure}[hpbt]
   \epsfig{figure=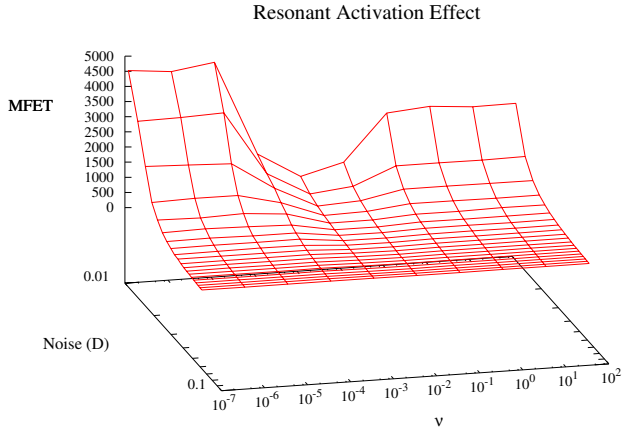, angle=-90, width=10cm}
\caption{\small Resonant Activation Effect due to the
competitive presence of noise and the periodic treatment.
The plot shows the mean escape time as a function of the
noise intensity $D$ and the frequency of the treatment $\nu$. The
parameters are the same than those of Fig.\ref{fig:pot}.}
 \label{fig:ra}
\end{figure}

Fig.~\ref{fig:mpt} reports also the behavior
of both the Kramers mean exit time and mean return time
calculated in the case of the fixed potential, the
expressions are the following:
 i) $\tau_{Exit}=2\pi e^{2h/D}/\sqrt{|U''(x_3)U''(x_2)|}$; ii) $\tau_{Ret} =
\tau_{Exit} + \tau_f$ with $\tau_f=2\pi e^{2h/D} /
\sqrt{|U''(x_1)U''(x_2)|}$, where $h$ is the height of the
barrier in absence of treatment.
The ratio $\tau_{Exit} / \tau_{Ret}$ shows only a constant
reference value because of the equal depth of the
potential.
Finally, in Fig.\ref{fig:mpt2} the behavior of the standard
deviation of the FET and FRT distributions is reported,
showing also a well visible minimum in the ratio
$\sigma_{E}/\sigma_{R}$ at $D \approx 0.015$, close to the
noise intensity at which the minimum of the ratio MFET/MFRT
has been observed. This means that also the relative precision of the exit/return times evaluations shows an optimal value in the same noise surroundings than the corresponding ratio of the means.
Moreover, the MFET has been
analyzed as a function of the frequency of the treatment,
founding the resonant activation
phenomenon \cite{ra} (see Fig.\ref{fig:ra}) in a certain range of the
noise intensity (see also \cite{Dybiec1,Dybiec2}). Note
that the numerical simulations, as discussed in this work,
have been performed with the frequency of treatment $\nu =
10^{-3}$, corresponding to the minimum of the MFET
detectable in the ($D,\nu$) space (Fig.\ref{fig:ra}).

\section{Conclusions}

In this paper we
investigate a mathematical model describing the growth of
tumor in the presence of the immune response of a host
organism. The model is supplemented with periodic treatment
and external fluctuations in the tumor growth rate. The
formulation of the model is based on a reaction scheme
\cite{Chaplain,GARAY,Lefever_Garay,Lefever_Horsthemke,Prigogine_Lefever,Horsthemke_Lefever,Ebeling,Gudowska1,Gudowska2},
representative of the catalytic Michaelis-Menten scenario.

The resulting phenomenological equation modelling
cell-mediated immune surveillance against cancer exhibits
bistability. The two stationary points correspond to the
state of a stable tumor and the state of its extinction.
The influence of therapeutic factors is introduced in the
form of an on-off switch of the cyclic treatment, performed
with a certain intensity and frequency, which (when
switched on) acts in favor of a decrease of the tumor cell
concentration. Under the influence of weak fluctuations,
the model is analyzed in terms of a stochastic differential
equation. It has the form of an overdamped Langevin-like
dynamics in the external quasi-potential represented by a
double well. We analyze properties of the system within the
range of parameters for which the potential wells are of
the same depth and when the periodic treatment alone (without external noise) is
insufficient to overcome the barrier height and to cause
cancer extinction.
\\
The aim of our study was to explore the dependence between
the intensity of fluctuations and the transition time from
the stable tumor to the extinction state. We performed a
series of numerical simulations of the stochastic
trajectories of the system and analyzed the distributions
of their duration times. Without addition of noise, the
concentration of cancer cells remains close to the stable
tumor state: the treatment is insufficient to cause the
extinction. A small amount of noise added to the system can
give rise to cancer extinction in the time intervals when
the treatment is switched on. At higher noise intensities,
the extinction occurs even in the "treatment-off"
intervals. However, the presence of noise also causes the
tumor recurrence. Cancer regrowth events occur during the
no-treatment time interval for low noise intensity and, at
higher noise intensity, may also be detected during the
"treatment-on" intervals. At low noise intensities, the
return probability can be even more than one order of magnitude
lower than the extinction probability, but at sufficiently
high noise the times to cross the barrier in both forward
and back direction are approximately equal.

We found that there exists an optimal noise intensity
$D_{opt}$ for which the ratio MFET/MFRT of the mean first
extinction time to the mean first return time is the
smallest. Since the duration of particular realizations of
extinction and recurrence processes vary in the presence of
noise, we also studied the width of FET and FRT
distributions. Interestingly enough, the ratio of the
corresponding standard deviations $\sigma_{E}/\sigma_{R}$
also has a minimum at a certain noise intensity $D_{\sigma
opt}$, which is close to $D_{opt}$. Not far from the same value
we also find a maximum for the 'extinction without return'
probability as a function of the noise intensity.
Our study shows that the presence of noise gives the
possibility of tumor extinction even at a weak radiation or
chemical treatment. Treatment strategies which take into
account the noise contribution can reduce the radiation (or
chemotherapeutic) dose. On the other hand, the presence of
noise implies also the possibility of a recurrence. We
show, however, that it is possible to observe an optimal
noise value in the competition between the extinction and
return events.

\vspace{0.5cm} This work was supported by MUR and
INFM-CNISM. A.F. acknowledges the Marie Curie TOK grant
under the COCOS project (6th EU Framework Programme,
contract No: MTKD-CT-2004-517186).
A.O-M. acknowledges the support by Volkswagen Foundation, project No. I/80424.

\end{document}